\documentclass{article} 
\usepackage[dvips]{graphics}

\newcommand{\sol}{M_{\odot}}

\newcommand{\rhoh}{\rho_{_H}}

\newcommand{\rhon}{\rho_{_N}}

\newcommand{\gammaT}{\widetilde{\gamma}}

\newcommand{\gammaH}{\widehat{\gamma}}
\newcommand{\phiH}{\widehat{\phi}}
\newcommand{\betaT}{\widetilde{\beta}}
\newcommand{\KT}{\widetilde{K}}

\newcommand{\DT}{\widetilde{D}}

\newcommand{\LaplaceT}{\widetilde{\triangle}}
\newcommand{\auz}{\alpha u^0}
\newcommand{\sgamma}{\sqrt{\gamma}}

\newcommand{\apj}{{\it Astrophys. J}}

\newcommand{\prl}{{\it Phys. Rev. Letters}}
\newcommand{\mnras}{{\it Mon. Not. Roy. Astro. Soc.}}
\newcommand{\ptp}{{\it Prog. Theor. Phys.}}
\newcommand{\prd}{{\it  Phys. Rev.}}
\newcommand{\aap}{{\it  Astron. Astrophys.}}

\begin{document}

\begin{center}
{\LARGE A Way to 3D Numerical Relativity} \medskip \\
{\Large --- Coalescing Binary Neutron Stars ---} \bigskip \\

{\large Takashi NAKAMURA \\
Yukawa Institute for Theoretical Physics\\
Kyoto University, Kyoto 606-01, Japan \bigskip \\
Ken-ich Oohara \\
Department of Physics, Niigata University \\
Niigata, 950-2181, Japan}
\end{center}

\section{Introduction}

Long baseline interferometers to detect gravitational waves such as
TAMA\-300 \cite{tama}, GEO600 \cite{geo}, VIRGO \cite{virgo}, LIGO
\cite{ligo} will be in operation by the end of this century or the
beginning of the 21st century. One of the most important sources of
gravitational waves for such detectors are coalescing binary neutron
stars. PSR1913+16 is the first binary neutron stars found by Hulse and
Taylor. The existence of the gravitational waves is confirmed by
analyzing the arrival time of radio pulses from PSR1913+16.  At
present there are three systems like PSR1913+16 for which coalescing
time due to the emission of gravitational waves is less than the age
of the universe. From these observed systems the event rate of the
coalescing binary neutron stars is estimated as $10^{-6}\sim
2\times10^{-5}$ events yr$^{-1}$ galaxy$^{-1}$
\cite{phin91,nara91,vdhe96}. From the formation theory of binary
neutron stars, the event rate is estimated as $ 2\times10^{-5}\sim
3\times 10^{-4} $ events yr$^{-1}$ galaxy$^{-1}$
\cite{tutu94,lipu97,yung98}. If the event rate is $ \sim
2\times10^{-5}$, two different methods yield the agreement. Since the
number density of galaxies is $\sim 10^{-2}$/Mpc$^3$, we may expect
several coalescing events/year within 200Mpc ($\sim 2\times
10^{-7}$events /Mpc$^3$). The coalescing binary neutron stars is  a
possible central engine of cosmological gamma ray bursts. Coalescing
binary neutron stars is also a possible site of r-process element
production.

The eccentricity $e$ and the semimajor axis $a$ of binary neutron
stars decrease due to the emission of gravitational waves, 
and these quantities are related with each other as
\begin{equation}
e=e_0 \left(\frac{a}{a_0}\right)^{\frac{19}{12}} 
\end{equation}
where $e_0$ and $a_0$ are the initial  eccentricity  and the semimajor
axis, respectively \cite{peter63,peter64}. If $a_0\sim R_\odot$ like
PSR1913+16, $e\sim 10^{-6}$ for $a\sim 500$km so that the orbit is
almost circular. The merging time $t_{\rm mrg}$ due to the emission of
the gravitational waves is given by
\begin{equation}
t_{\rm mrg} = 3{\rm min}\left( \frac{m_1}{1.4\rm M_\odot}\right)^{-1}
\left( {m_2 \over 1.4\rm M_\odot}\right)^{-1}
\left( {m_1+m_2 \over 2.8\rm M_\odot}\right)^{-1}
\left( {a \over 470{\rm km}}\right)^4,
\end{equation}
where $m_1$ and $m_2$ are each mass of neutron stars. The frequency
of the gravitational waves $\nu_{_{\rm GW}}$ is given by 
\begin{equation}
\nu_{_{\rm GW}} =  19 {\rm Hz}~
\left( {m_1+m_2 \over 2.8\rm M_\odot}\right)^{1\over2}
\left( {a \over 470{\rm km}} \right)^{-{3\over2}}
\end{equation}
This is just the frequency range ($\nu_{_{\rm GW}}=10{\rm Hz}\sim
1{\rm kHz}$) of long baseline interferometers to detect gravitational
waves such as  TAMA300,   GEO600, VIRGO, LIGO. The number of rotation
before the final merge is given by
\begin{equation}
N = 2635~ \left( {m_1 \over 1.4\rm M_\odot} \right)^{-1}
\left( {m_2 \over 1.4\rm M_\odot} \right)^{-1}
 \left({ m_1+m_2 \over 2.8\rm M_\odot}\right)^{-{1\over 2}}
 \left({ a \over 470{\rm km}} \right)^{5\over 2} 
\end{equation}

The merging phase of coalescing binary neutron stars is divided into
two stages; 1)The last three minutes \cite{cutl93} and 2) The last 
three milliseconds. In the first stage $a$ is much larger than the
neutron star radius $R_{\rm NS}$ and the rotation velocity of the
binary $v_r$ is $\sim 0.05c$ so that each neutron star can be
considered as a point particle. The post-Newtonian expansion will
converge in the first stage. However the quite accurate theoretical
calculations of the wave form are needed since only the uncertainty of
1/2635 in the theoretical prediction of the energy loss will cause one
rotation ambiguity at last. If the accurate theoretical template is
obtained \cite{blan96}, by  making a cross correlation with the
observational data we may determine each mass, spin, inclination and
the distance of binary neutron stars \cite{cutl94}. 
 
In the second stage $a$ is comparable to $ R_{\rm NS}$ and the gravity
is so strong that  the post-Newtonian expansion is not a good
approximation and the finite size effect of  neutron stars is
important. In the final collision phase, the general relativistic
hydrodynamics are also important. Only 3D numerical relativity can
study this important final merging phase related to gravitational
waves, gamma ray burst and production of r-process elements.

\section{Newtonian Simulations}

 To clarify the phenomena in the last three milliseconds many
 numerical simulations have been done so far. They are classified into
 several categories.
\medskip

\noindent
{\bf A)  Newtonian hydrodynamics without radiation reaction}
\medskip

In this category, using the Newtonian hydrodynamics, equilibria and
stability of the binary neutron stars are studied for various initial 
parameters and equation of states
\cite{on89,no89,rasio92,cent93,rasio94,houser94,new97}. 
\medskip

\noindent
{\bf B) A)+ simple radiation reaction up to contact of neutron stars}
\medskip
 
In this category, radiation reaction is included so that the binary
spirals in due to the emission of the gravitational waves. However
after the contact the radiation reaction is switched off
 \cite{davies94,zhuge94,zhuge96}.
\medskip

\noindent
{\bf C) A) + radiation reaction}
\medskip

In this category the radiation reaction is fully taken into account
even after the contact. However the computation is time consuming
because to estimate the radiation reaction potential two more Poisson
equations  other than the Poisson equation for the Newtonian
gravitational potential should be solved
 \cite{no89,on90,no91,sno92,sno93,rjs96,rjts97,rrj97}.

\begin{figure}[tbp]
  \begin{center}
    \leavevmode
    \scalebox{.65}{\includegraphics{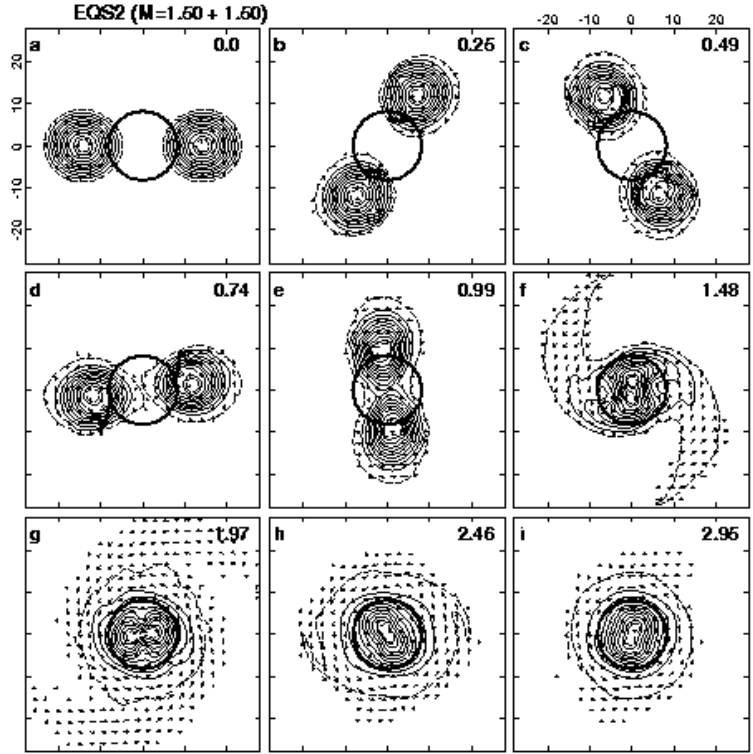}}
    \caption{ Density and velocity on the $x$-$y$ plane for EQSP2. The
      time in units of milliseconds is shown. Arrows indicate the
      velocity vectors of the matter. A fat line shows a circle of
      radius $2G M_t /c^2$.} \label{fig:eqsp2-density}
  \end{center}
\end{figure}
\begin{figure}[tbp]
  \begin{center}
    \leavevmode
    \scalebox{.5}{\includegraphics{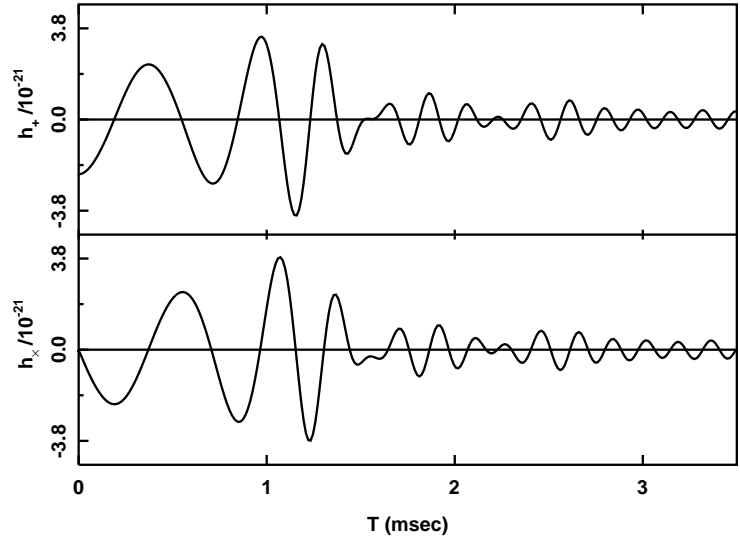}}
    \caption{Wave forms of $h_{+}$ and $h_{\times}$ observed on the
      $z$-axis at 10Mpc for EQSP2.} \label{fig:eqsp2-wave}
  \end{center}
\end{figure}

 In Fig.\ref{fig:eqsp2-density} we show an typical example of such a 
simulation \cite{on90}. In this case each mass of the binary is the
same. We start the simulation from the equilibrium obtained in
category {\bf A)}. Due to the loss of the angular momentum by
gravitational waves the merging starts spontaneously. As merging
proceeds, two arm spiral  extends outward. The spiral arms become
tightly bound later and becomes a disk. The final result is an  almost
axially symmetric central object and a disk around it. In
Fig.\ref{fig:eqsp2-density} the thick  circle shows the Schwarzschild
radius of the total mass. Although it is not possible to say the
formation of a black hole in Newtonian simulations, we expect that the
central object is a black hole. So the final destiny may be a black
hole and a disk. This behavior is seen also in many other simulations
in category {\bf B)} and {\bf C)}. In Fig.\ref{fig:eqsp2-wave} we show
a typical wave pattern of the gravitational waves.

\begin{figure}[tbp]
\begin{center}
  \leavevmode
  \scalebox{.65}{\includegraphics{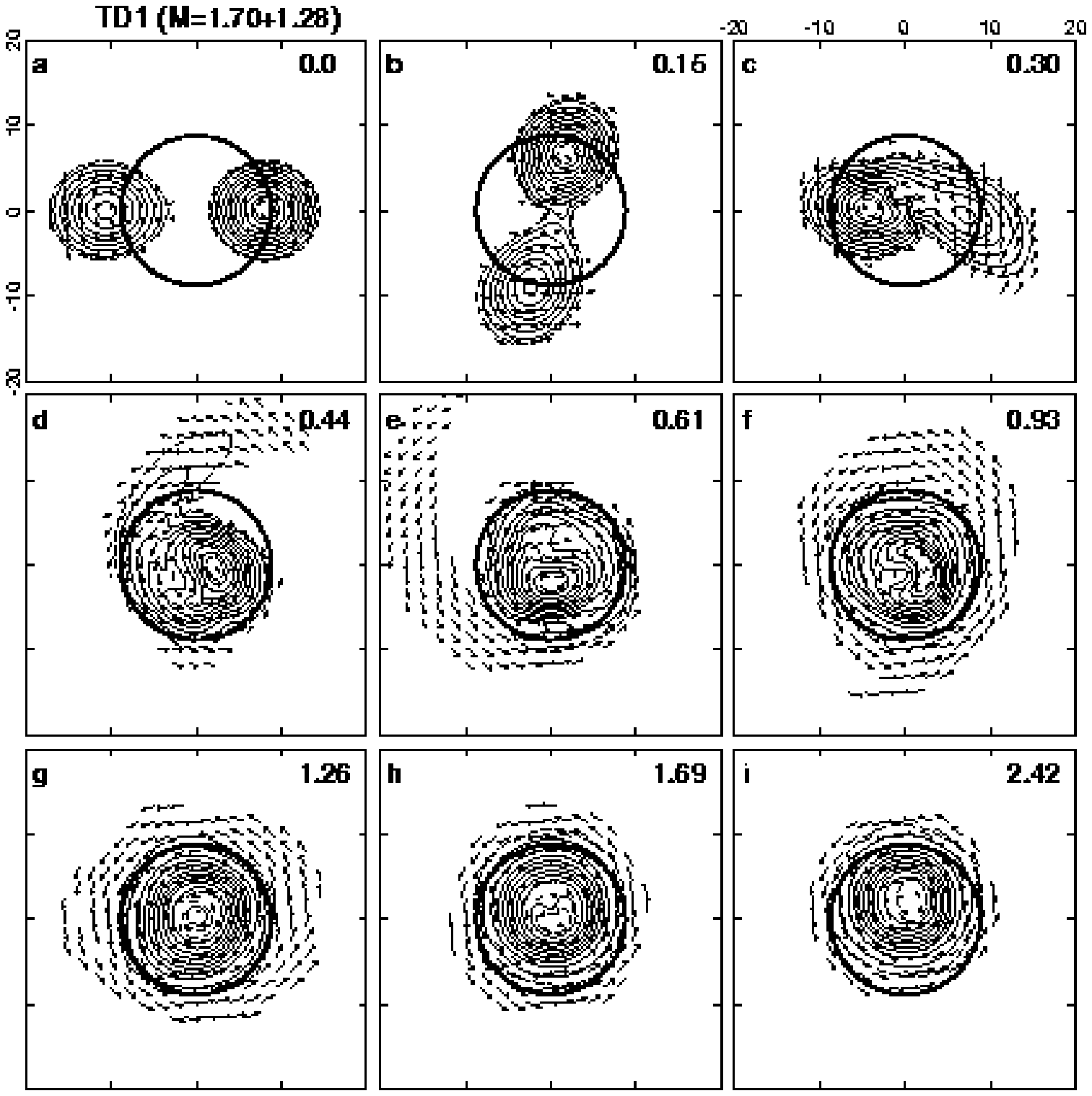}}
  \caption{Density and velocity on the $x$-$y$ plane for TD1.}
  \label{fig:td1-density}
\end{center}
\end{figure}

In Fig.\ref{fig:td1-density}, a simulation for different mass case is
shown\cite{no91}. In this case the lighter neutron star (left in
Fig.\ref{fig:td1-density}-a) is tidally disrupted by the heavier
neutron star. We see only one arm . However the final object is more
or less similar to the equal mass case, that is, a black hole and a
disk. This final object may be relevant to the central engine of gamma
ray burst. Ruffert will discuss in this volume on the coalescing
binary neutron stars and gamma ray bursts using their simulations
including the emission of neutrinos\cite{rjs96,rjts97}.

\medskip
\noindent
{\bf D) A)+1PN effect of Post Newtonian Hydrodynamics and Radiation
  Reaction}
\medskip
  
\begin{figure}[tbp]
  \begin{minipage}{.49\textwidth}
  \begin{center}
    \leavevmode
    \scalebox{.5}{\includegraphics{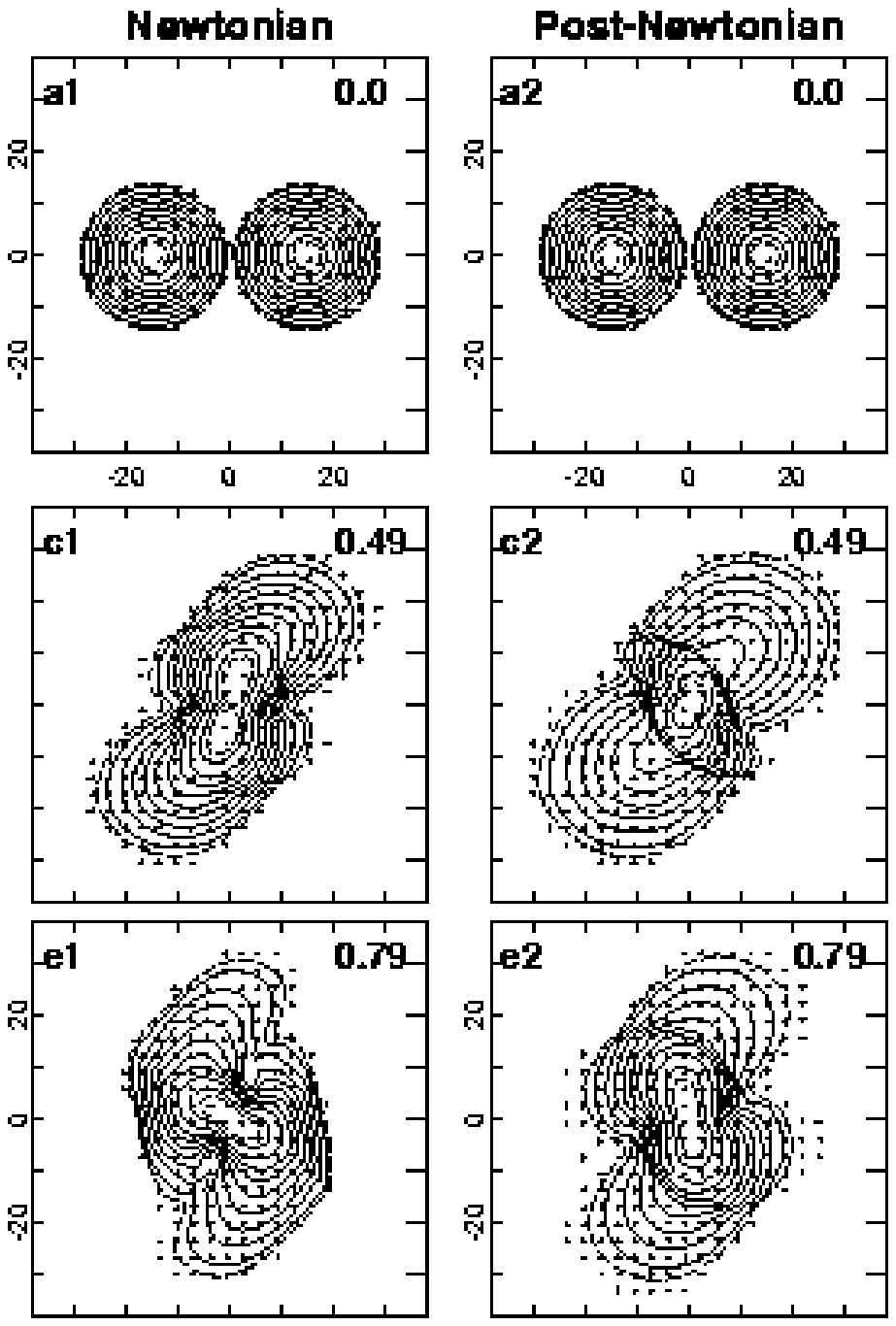}}
  \end{center}
  \end{minipage}
  \begin{minipage}{.49\textwidth}
  \begin{center}
    \leavevmode
    \scalebox{.5}{\includegraphics{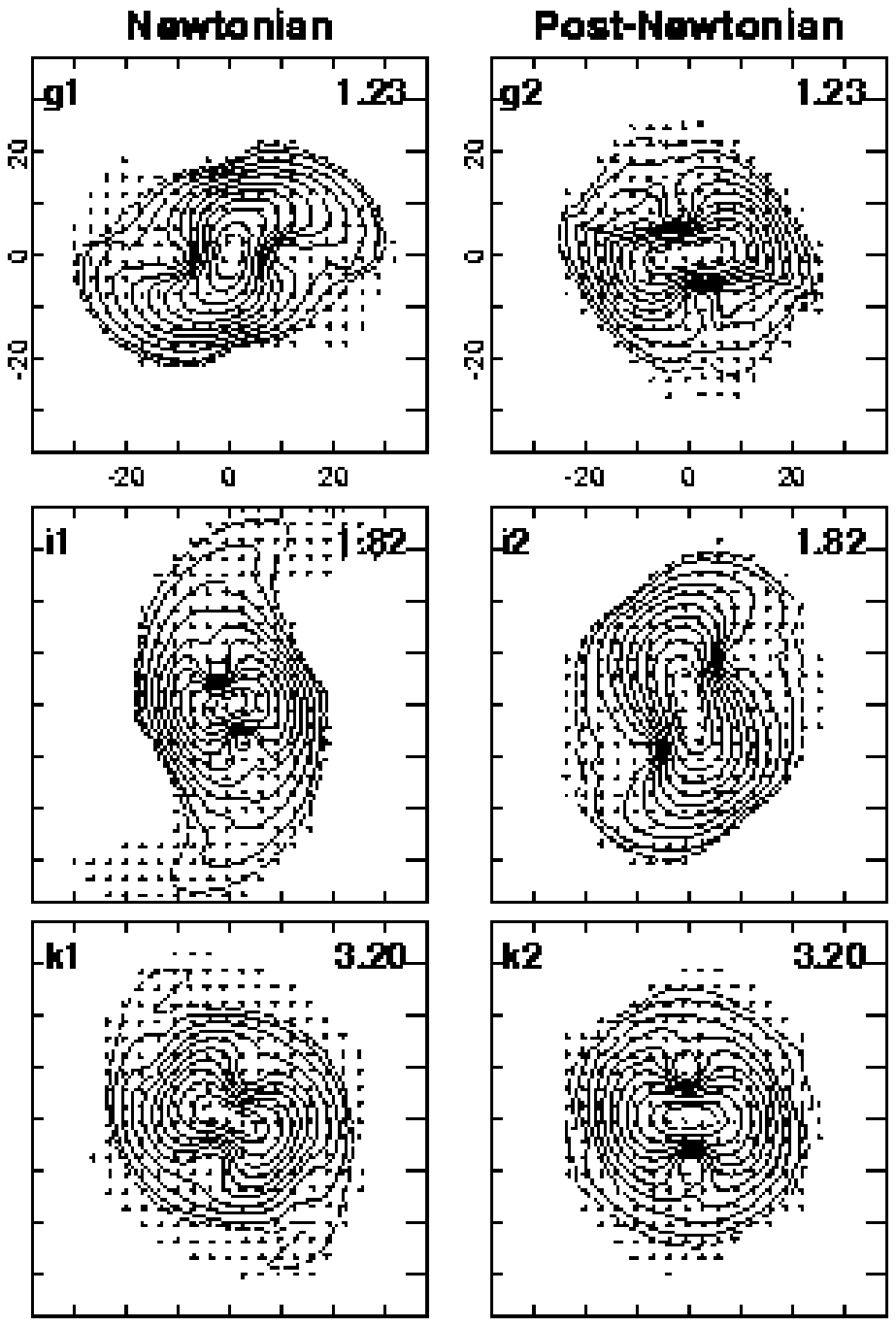}}
  \end{center}
  \end{minipage}
    \caption{Density and velocity on the $x$-$y$ plane. The left and
      right figures are for the Newtonian (N) and post-Newtonian (PN)
      calculations, respectively. Notations are the same as for
      Fig.\protect\ref{fig:eqsp2-density}.}
    \label{fig:pndn-density}
\end{figure}

\begin{figure}[tbp]
  \begin{center}
    \leavevmode
    \scalebox{.4}{\includegraphics{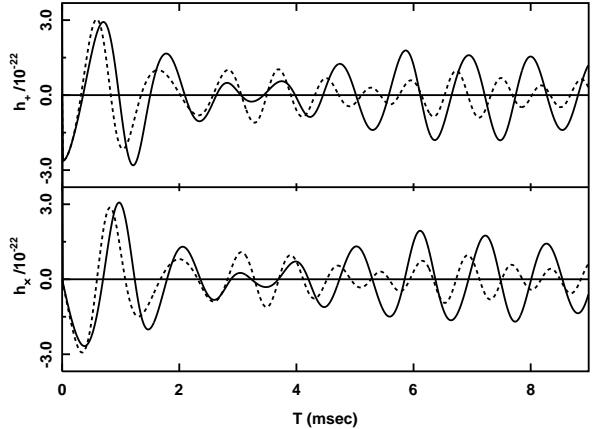}}
    \caption{Wave forms of $h_{+}$ and $h_{\times}$ observed on the
      $z$-axis at 10Mpc. The solid and dashed lines are for PN and N,
      respectively.}  \label{fig:pndn-wavef}
  \end{center}
\end{figure}

Since neutron stars are general relativistic compact objects, the
Newtonian hydrodynamics  is zeroth approximation to the problem. One
possible refinement is to include 1PN force [i.e. $O \left((v/c)^2
\right)$ effect of the general relativistic gravity after the
Newtonian gravity]. Oohara and Nakamura performed such a simulation
\cite{on92}. They first include 1PN force to the simulation in
Fig.\ref{fig:eqsp2-density} and found that  the 1PN effect is too
strong. They decrease the mass of the neutron star to $0.62M_\odot$
from$1.5M_\odot$  keeping the radius of the neutron the same  so that
the gravity becomes weaker. To see the difference between Newtonian
and 1PN cases, in Fig.\ref{fig:pndn-density} two simulations from the
same initial data are shown \cite{on92}. We clearly found that they
are different even in this rather weak gravity case. In
Fig.\ref{fig:pndn-wavef} the wave forms are shown. The solid line
(1PN) and the dashed line (Newton) are different. Due to the strong
gravity effect of 1PN force, the strong shock  in the central is
formed and the coalescence is delayed. If  the higher  post Newtonian
effects (2PN,...) are included, different results may be
obtained. This suggests  that fully general relativistic simulations
are needed. This suggests also that  all the results of coalescing
binary neutron stars based on Newtonian hydrodynamics or
Post-Newtonian hydrodynamics need to be viewed with caution.

There is another point to be considered. Suppose the point particle
limit of each neutron star. In the Newtonian gravity there exists
a circular orbit for any small separation $a$. However in general
relativity a stable circular orbit does not exist for $a< r_{\rm
ISCO}$ where ISCO stands for Innermost Stable Circular Orbit. For a
test particle motion in the Schwarzschild black hole $r_{\rm
ISCO}=6GM/c^2$ where $M$ is the mass of black hole. For equal mass
point mass binary $r_{\rm ISCO}$ is estimated as $\sim 14Gm/c^2$ where
$m$ is each mass of the binary \cite{kidd92}. Since the radius of the
neutron star is $\sim 5Gm/c^2$, when two neutron stars contact, the
separation $a$ is $\sim 10Gm/c^2$, which is comparable to $r_{\rm
  ISCO}$ in the point particle limit. Therefore in the last three
milliseconds the general relativity is also important in the orbital
motion. Recently ISCO for binary neutron stars including the effect of
general relativity and the finite size of the neutron stars has been
extensively studied. Shibata discusses ISCO problem in this volume. 

\section{General Relativistic Simulations}

For full general relativistic 3D simulations of coalescing binary
neutron stars, we need ISCO of binary neutron stars as initial data.
However at present such initial data are not available. We therefore
discuss the development of 3D general relativistic code and test
simulations. We present only the summary of the basic equations.
For details of formalisms, basic equations and numerical methods
please refer \cite{ons97}.

\subsection{Basic Equations}

\subsubsection{Initial Value Equations}

We adopt the (3+1)-formalism of the Einstein equation. Then the line
element is written as
\begin{equation}
  ds^2 = - \alpha^2 dt^2 + \gamma_{ij} ( dx^i + \beta^i dt )
  ( dx^j + \beta^j dt),
\end{equation}
where $\alpha$, $\beta^i$ and $\gamma_{ij}$ are the lapse function,
the shift vector and the  metric of 3-space, respectively.
The lapse and shift represent the coordinate degree of freedom and its
choice is essential for numerical simulations.  $\gamma_{ij}$ is
dynamical variable including  gravitational waves. 

We assume the perfect fluid. The  energy momentum tensor is given by
\begin{equation}
  \label{eq:enmom}
  T_{\mu \nu} = ( \rho + \rho \varepsilon + p ) u_{\mu} u_{\nu}
  + p g_{\mu \nu},
\end{equation}
where $\rho$, $\varepsilon$ and $p$ are the proper mass density, the
specific internal energy and the pressure, respectively, and $u_\mu$
is the four-velocity of the fluid. We define
$ \rhoh \equiv n^\mu n^\nu T_{\mu \nu}, \ \
  J_i \equiv - h_i{}^\mu n^\nu T_{\mu \nu} \ \ \mbox{and} \ \ 
  S_{ij} \equiv h_i{}^\mu h_j{}^\nu T_{\mu \nu}$
where $n_\mu$ is the unit  normal to the $t$=constant hypersurface and
$h_{\mu \nu}= g_{\mu \nu} + n_\mu n_\nu$ is the projection tensor to
the hypersurface.

The initial data should satisfy the Hamiltonian and momentum 
constraint equations given by
\begin{eqnarray}
  & & R + K^2 - K_{ij} K^{ij} = 16 \pi \rhoh , \label{eq:hconst} \\
  & & D_j \left( K^j{}_i - \delta^j{}_i K \right) = 8 \pi J_i,
      \label{eq:mconst}
\end{eqnarray}
where $R$ is the 3-dimensional Ricci scalar curvature, $K_{ij}$ is the
extrinsic curvature and $D_i$ denotes the covariant derivative with
respect to $\gamma_{ij}$.

\subsubsection{Relativistic Hydrodynamics}
In order to obtain equations similar to the Newtonian hydrodynamics
equations, we define $\rhon = \sgamma \auz \rho$ and $u_i^N =
{J_i}/{\auz \rho}$ where $\gamma = \mbox{det} (\gamma_{ij})$. Then the
equation for the conservation of baryon number is expressed as
\begin{equation}
  \label{eq:hydrob}
  \partial_t \rhon +   \partial_\ell \left( \rhon V^\ell \right) = 0 ,
\end{equation}
where $V^\ell = u^\ell/u^0$. The equation for momentum  is
{\arraycolsep = 2pt
\begin{eqnarray}
  \partial_t (\rhon u_i^N) +
  \partial_\ell \left( \rhon u_i^N V^\ell \right)
  & = & - \sgamma \alpha \partial_i p - \sgamma ( p + \rhoh )
  \partial_i \alpha \nonumber \\
  & &  + \frac{\sgamma \alpha J^k J^\ell}{2(p + \rhoh)} \partial_i
  \gamma_{k \ell} + \sgamma J_\ell \partial_i \beta^\ell .
  \label{eq:hydrom}
\end{eqnarray}
}%
The equation for internal energy  becomes
\begin{equation}
  \label{eq:hydroe}
  \partial_t (\rhon \varepsilon) +
  \partial_\ell \left( \rhon \varepsilon V^\ell \right) =
  - p \partial _\nu \left( \sgamma \alpha u^\nu \right).
\end{equation}
To complete hydrodynamics equations, we need an equation of state,
$p = p(\varepsilon, \rho)$

\subsubsection{Time Evolution of the Metric Tensor}

The evolution equation for the extrinsic curvature which is 
essentially the time derivative of $ \gamma_{ij}$ is given by
\begin{eqnarray}
  \label{eq:evolks1}
\partial_t K_{ij} & = & \alpha \left\{ R_{ij}
    - 8 \pi \left[ S_{ij} + {\textstyle \frac{1}{2}} \gamma_{ij}
      \left( \rhoh - S^\ell{}_\ell \right) \right] \right\}
  - D_i D_j \alpha  \\[.25em]
   & + & \alpha \left( K K_{ij}
    - 2 K_{i \ell} K^\ell{}_j \right),  \\[.25em]
   & + & K_{mi} \partial_j \beta^m
  + K_{mj} \partial_i  \beta^m - K_{ij} \partial_m \beta^m,
  \label{eq:SKbeta}
\end{eqnarray}
where $R_{ij}$ is the 3-dimensional Ricci tensor.

Now we define the conformal factor $\phi$ as
$  \phi \equiv \left( \mbox{det} \left[ \gamma_{ij} \right]
  \right)^{{1}/{12}}$
and $\gammaT_{ij}$ as $\gammaT_{ij} =  \gamma_{ij}/\phi^{4}$.
The conformal factor $\phi$ is determined by
\begin{equation}
  \label{eq:conf-t}
  \LaplaceT \phi = - \frac{\phi^5}{8} \left( 16 \pi \rhoh
    + K_{ij} K^{ij} - K^2 - \phi^{-4} \widetilde{R} \right),
\end{equation}
where $\LaplaceT$ and $\widetilde{R}$ is the Laplacian and the scalar
curvature with respect to $\gammaT_{ij}$, respectively. The time
evolution of $\gammaT_{ij}$ is given by
\begin{eqnarray}
  \label{eq:gammat}
  \partial_t \gammaT_{ij} & = & - 2 \alpha \left( \KT_{ij} -
    \frac{1}{3} \gammaT_{ij} K \right)
   + \DT_i \betaT_j + \DT_j \betaT_i
    - \frac{2}{3} \gammaT_{ij} \DT_\ell \beta^\ell \\
  & \equiv & A_{ij}^T, \nonumber
\end{eqnarray}
where
\begin{equation}
  \label{eq:KTevol}
  \KT_{ij} = \phi^{-4} K_{ij}, \quad \betaT_{i} = \phi^{-4} \beta_i
\end{equation}
and $\DT_i$ denotes the covariant derivative with respect to
$\gammaT_{ij}$.

\subsubsection{Coordinate Conditions}
\label{sec:coordinate}

On of the most serious technical problems in numerical relativity is the
control of the space and time coordinates. Difficulties arise from (a) 
coordinate singularities caused by strong gravity and the dragging
effect, where coordinate lines may be driven towards or away from each
other, and (b) spacetime singularities which appear when a black hole
is formed. We should demand coordinate conditions so that these
singularities can be avoided.
\medskip

\noindent
\underline{\em Spatial coordinates --- the shift vector $\beta^i$}
\medskip

We demand  $  \partial_i A_{ij}^T = 0$. Then $A_{ij}^T$ becomes
transverse-traceless. Equation (\ref{eq:gammat}) is reduced to the
equation for the shift vector $\beta^i$ as
\begin{eqnarray}
  \lefteqn{\nabla^2 \beta^i + \frac{1}{3} \partial_i
    \left( \partial_\ell \beta^\ell \right) = } \nonumber \\
  & & \partial_j \left[ 2 \alpha \left( \KT_{ij} - \frac{1}{3}
      \gammaT_{ij} K \right) \right] \label{eq:beta} \\
  & - & \partial_j \left[ \gammaH_{j \ell} \partial_i \beta^\ell
    + \gammaH_{i \ell} \partial_j \beta^\ell
    - \frac{2}{3} \gammaH_{ij} \partial_\ell \beta^\ell
    + \beta^\ell \partial_\ell \gammaH_{ij} \right], \nonumber
\end{eqnarray}
where $\nabla^2$ is the simple flat-space Laplacian and $\gammaH_{ij}
= \gammaT_{ij} - \delta_{ij}$.  
\medskip

\noindent
\underline{\em Time Slicing --- the lapse function $\alpha$}
\medskip

We require a time slice to have the following properties: (1) In the
central region, spacetime singularities should be avoided. These
singularities are regions of spacetime where curvature, density and
other quantities are infinite. (2) In the exterior vacuum region, the
metric should be stationary except for the wave parts to catch
gravitational waves numerically.

The lapse function $\alpha$ is given by
\begin{equation}
  \label{eq:confalpha}
  \alpha = \exp \left[ -2 \left( \phiH +
      \frac{\phiH^3}{3} + \frac{\phiH^5}{5} \right)
  \right] ,
\end{equation}
where $\phiH = \phi - 1$. In this  slicing, the space outside the
central matter quickly approaches the Schwarzschild metric. Since the
non-wave parts decrease as $O(r^{-4})$ for large $r$, we can catch
gravitational waves numerically with $O\left((M/r_c)^3\right)$
accuracy, where $M$ is mass of the system and $r_c$ is distance from
the origin of the position where gravitational waves are evaluated. 

\section{Coalescing Binary Neutron Stars in 3D Numerical Relativity}

We show our test simulations of coalescing binary neutron stars. We
place  two spherical neutron stars of mass $M = 1.0\sol$ and radius
$r_0 = 6M$ with density distribution ($\rhon$) of $\gamma = 2$
polytrope at $x = 7.2M$ and $x =  -7.2M$. We add a rigid rotation with
angular velocity $\Omega$ as well as an approaching velocity to this
system such that
\begin{eqnarray}
  v_x^N & = & \left\{
    \begin{array}[c]{ll}
      - \Omega y - \Omega r_0 \ & \mbox{if} \ x > 0, \\[.5em]
      - \Omega y + \Omega r_0 \ & \mbox{if} \ x < 0,
    \end{array} \right. \\[.5em]
  v_y^N & = & \Omega x.
\end{eqnarray}

We  adopt $(x, y, z)$ coordinates with 201$\times$201$\times$201 grid
size (9GB) and  typical CPU time being 10 hours on VPP300(FUJITU).
The details of a numerical method is shown in \cite{ons97}.

\begin{figure}[tbp]
  \begin{minipage}[t]{.45\textwidth}
  \begin{minipage}[t]{.48\textwidth}
    \begin{center}
    \leavevmode
    \scalebox{.2}{\includegraphics{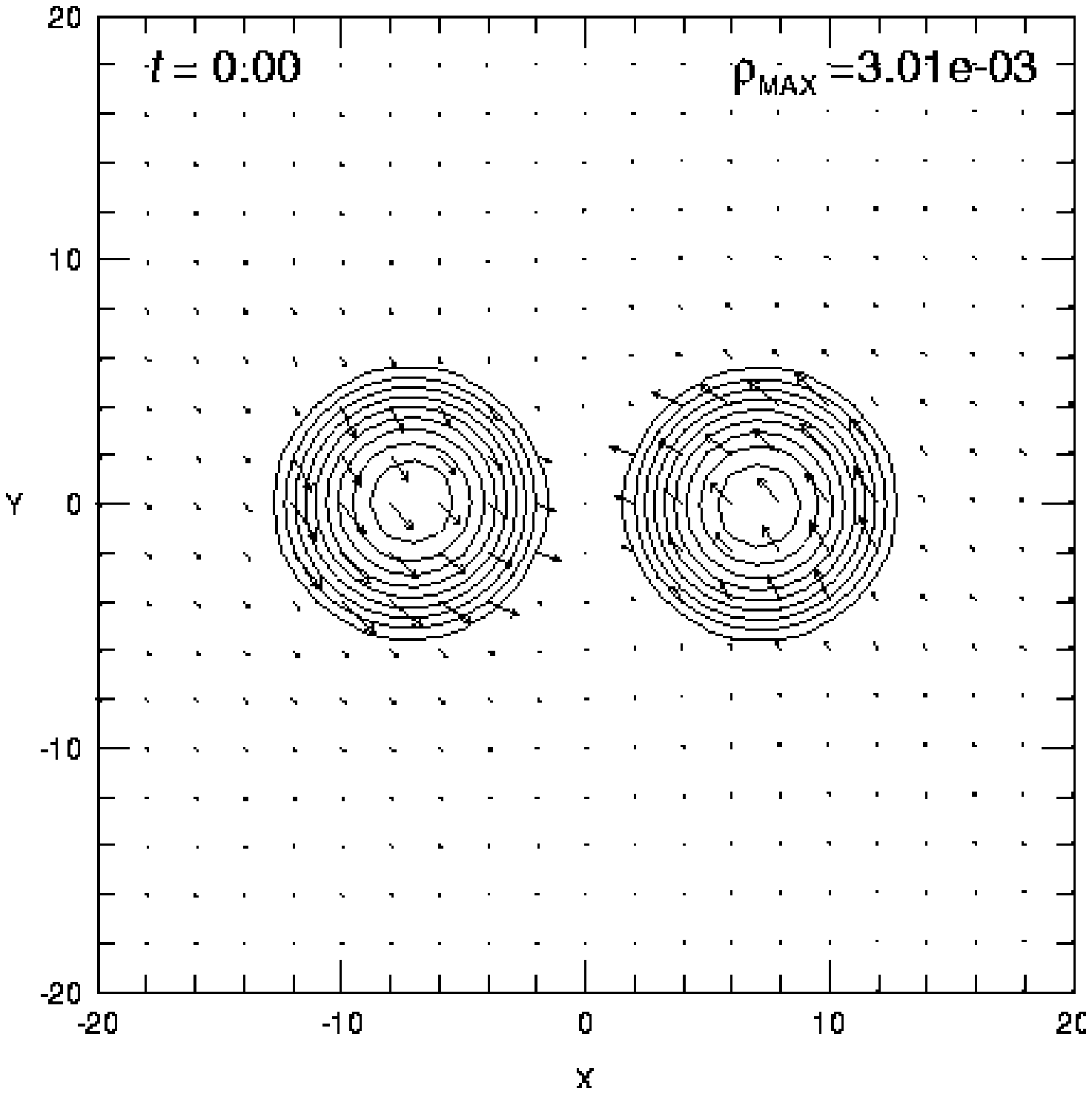}}
    \end{center}
  \end{minipage}
  \hfill
  \begin{minipage}[t]{.48\textwidth}
    \begin{center}
    \leavevmode
    \scalebox{.2}{\includegraphics{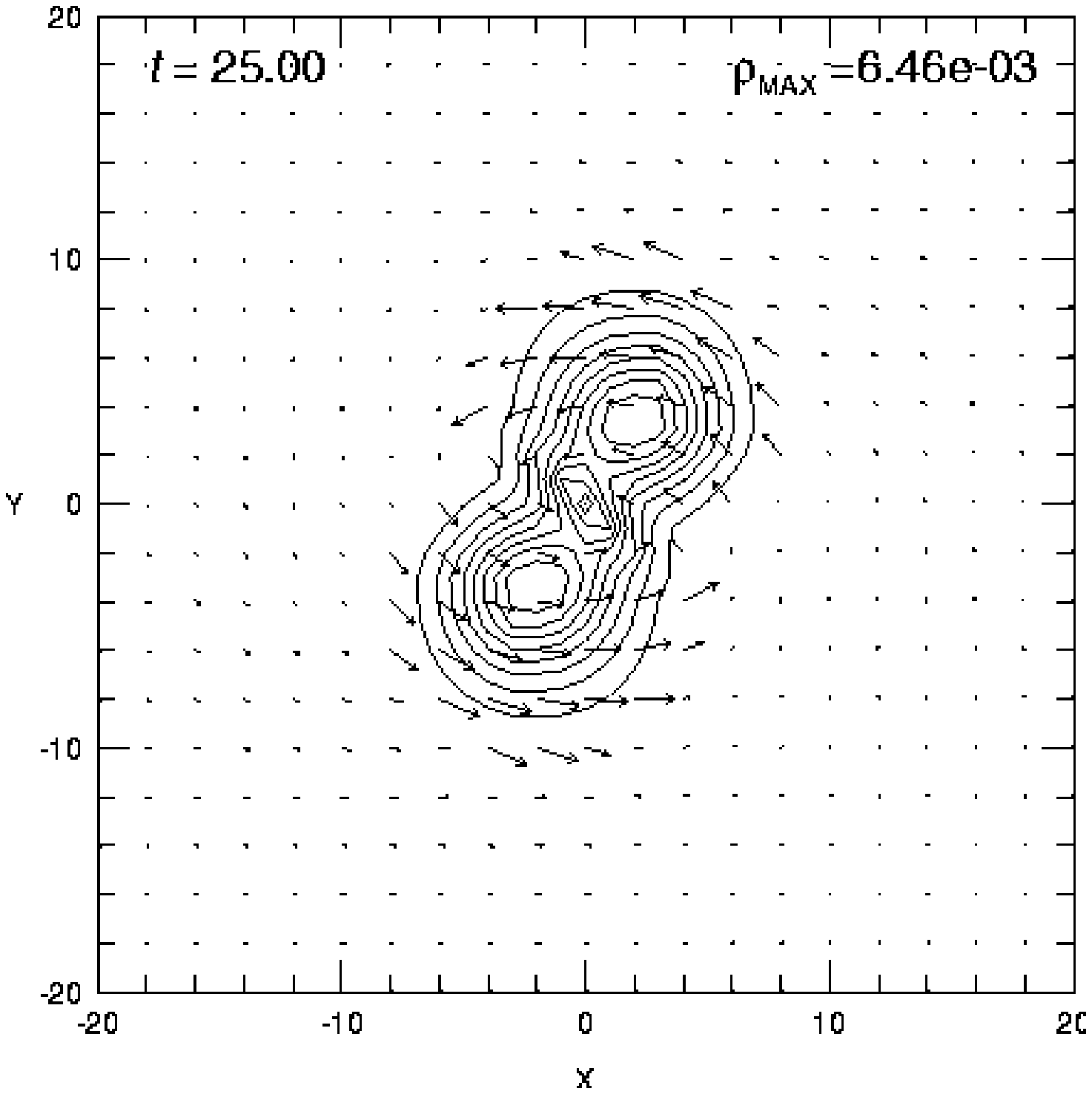}}
    \end{center}
  \end{minipage}
  \hfill
  \begin{minipage}[t]{.48\textwidth}
    \begin{center}
    \leavevmode
    \scalebox{.2}{\includegraphics{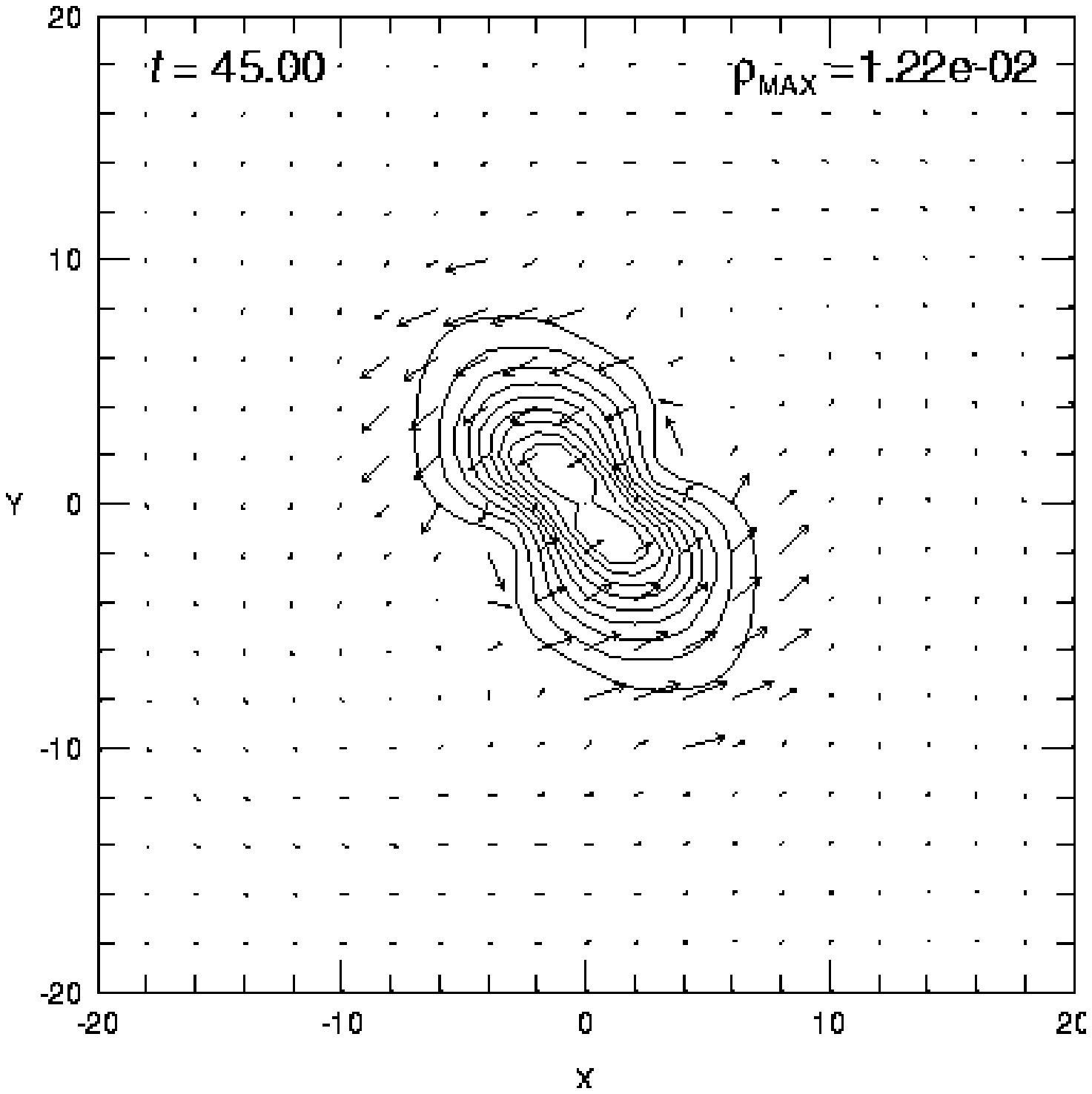}}
    \end{center}
  \end{minipage}
  \hfill
  \begin{minipage}[t]{.48\textwidth}
    \begin{center}
    \leavevmode
    \scalebox{.2}{\includegraphics{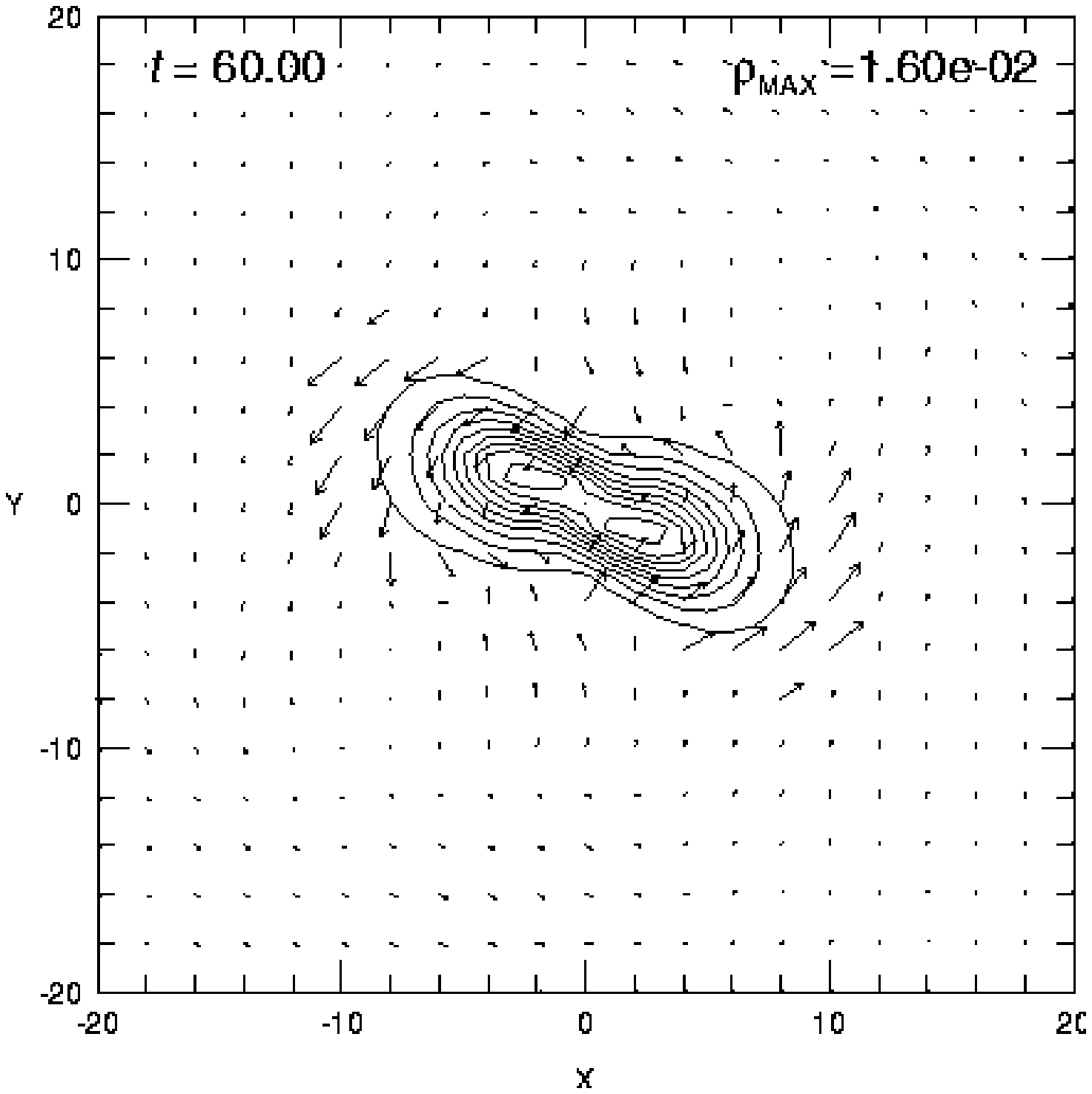}}
    \end{center}
  \end{minipage}
    \caption{Mass density $\rho_N$ on the $x-y$ plane.
      Arrows indicate the velocity vector $V^i$.
      Quantities are represented in units of $c = G = M_\odot = 1$.}
    \label{fig:bincol-density}
  \end{minipage}
  \hfill
  \begin{minipage}[t]{.45\textwidth}
  \begin{minipage}[t]{.48\textwidth}
    \begin{center}
    \leavevmode
    \scalebox{.2}{\includegraphics{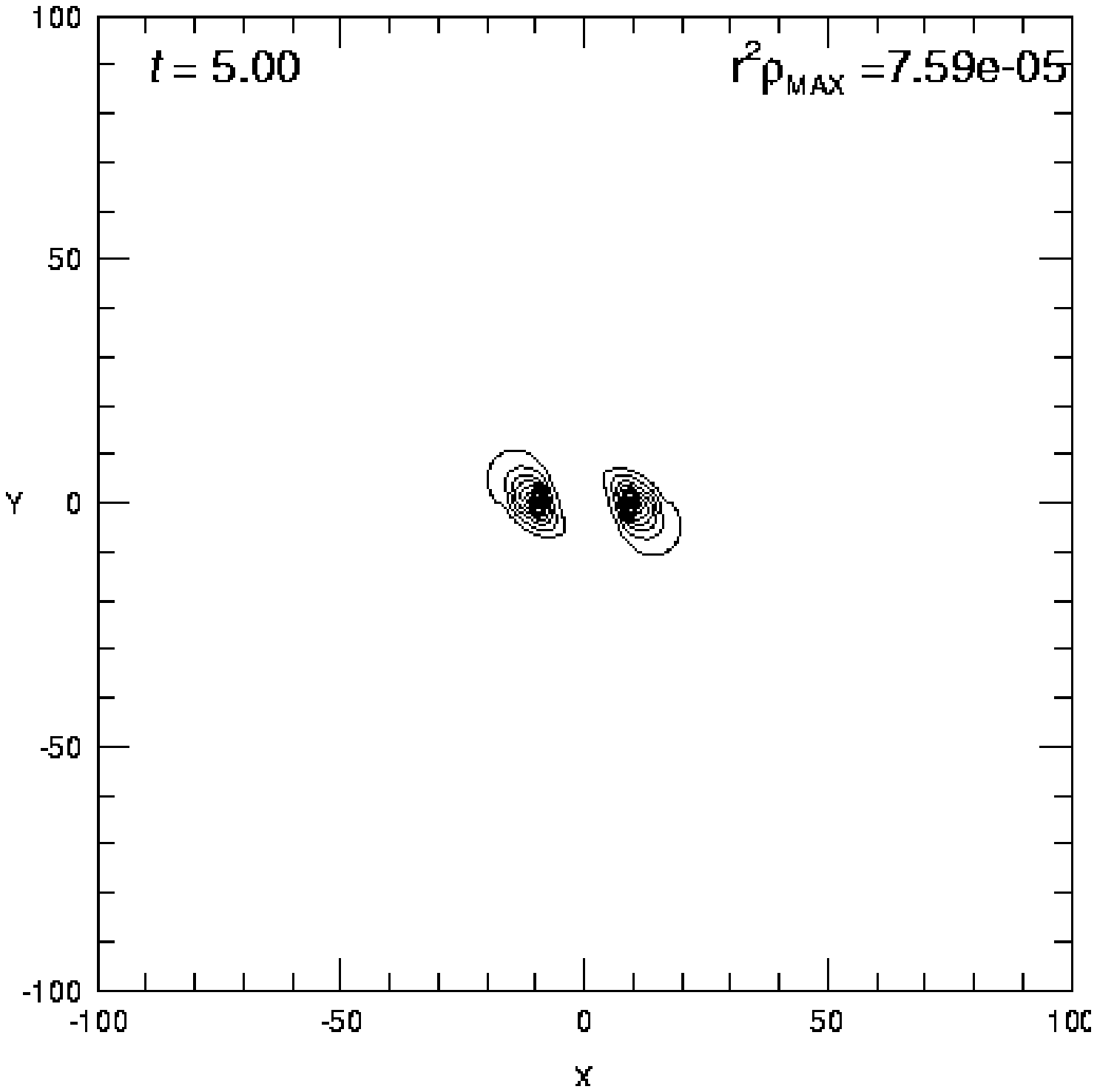}}
    \end{center}
  \end{minipage}
  \hfill
  \begin{minipage}[t]{.48\textwidth}
    \begin{center}
    \leavevmode
    \scalebox{.2}{\includegraphics{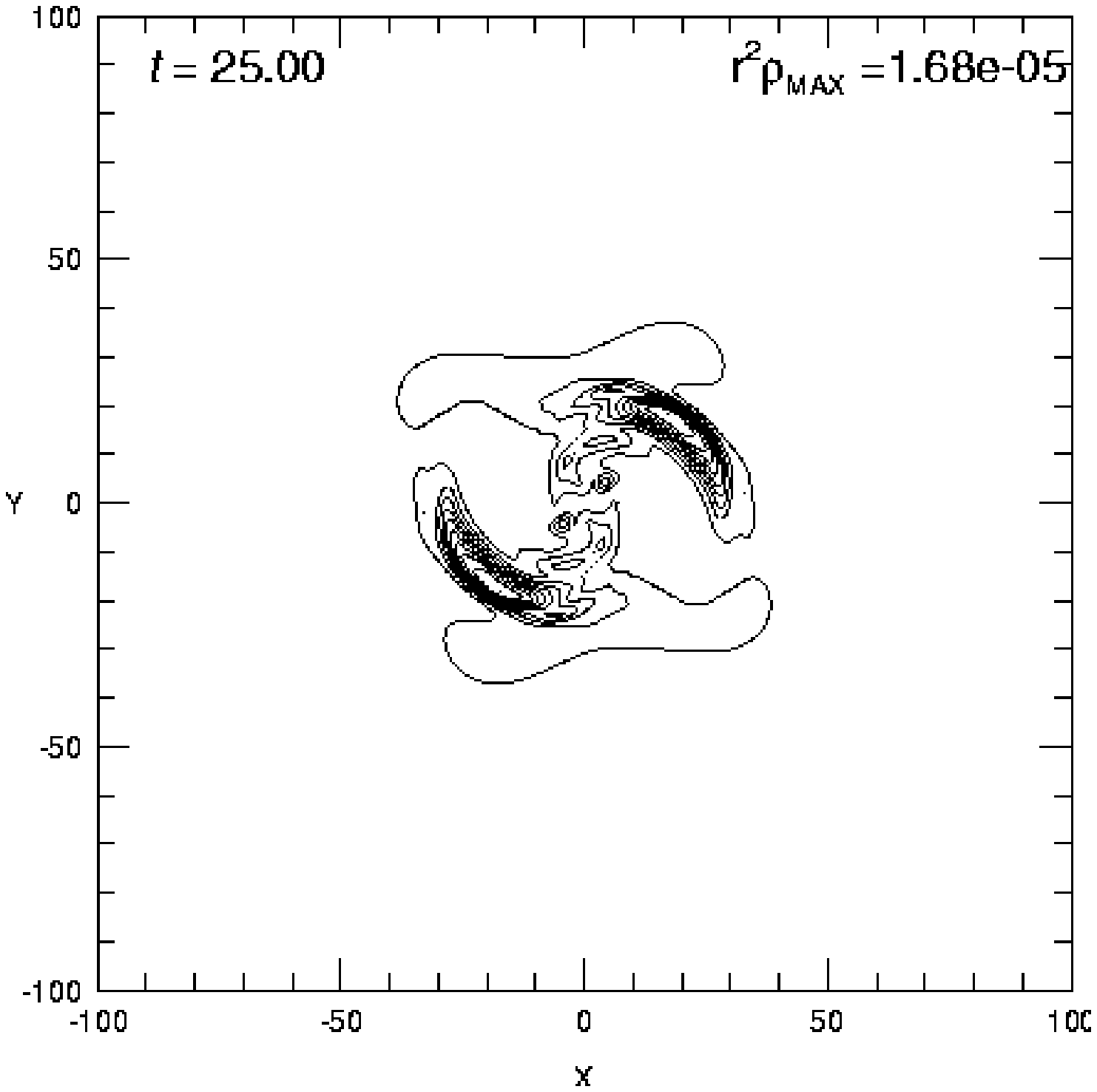}}
    \end{center}
  \end{minipage}
  \hfill
  \begin{minipage}[t]{.48\textwidth}
    \begin{center}
    \leavevmode
    \scalebox{.2}{\includegraphics{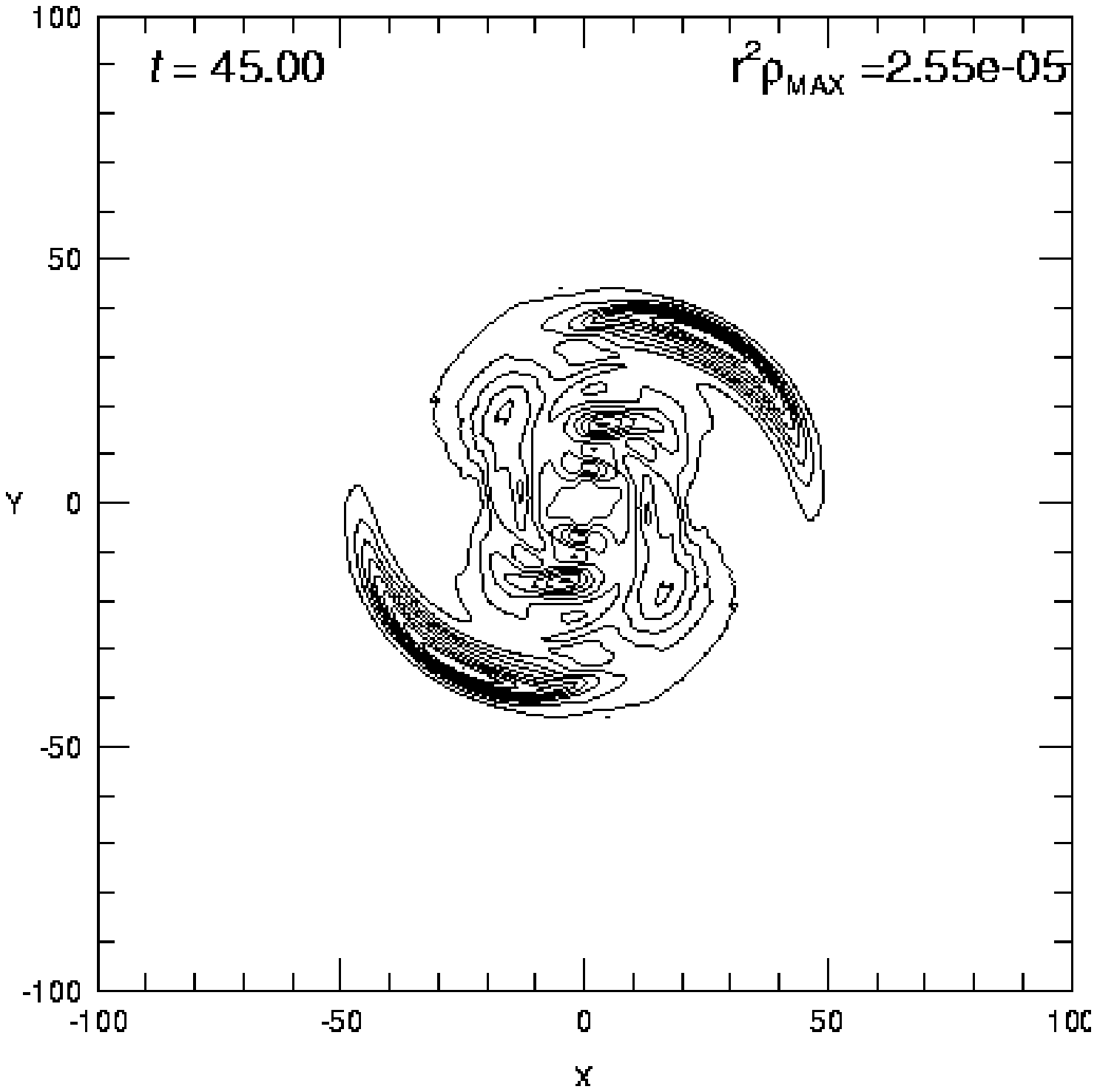}}
    \end{center}
  \end{minipage}
  \hfill
  \begin{minipage}[t]{.48\textwidth}
    \begin{center}
    \leavevmode
    \scalebox{.2}{\includegraphics{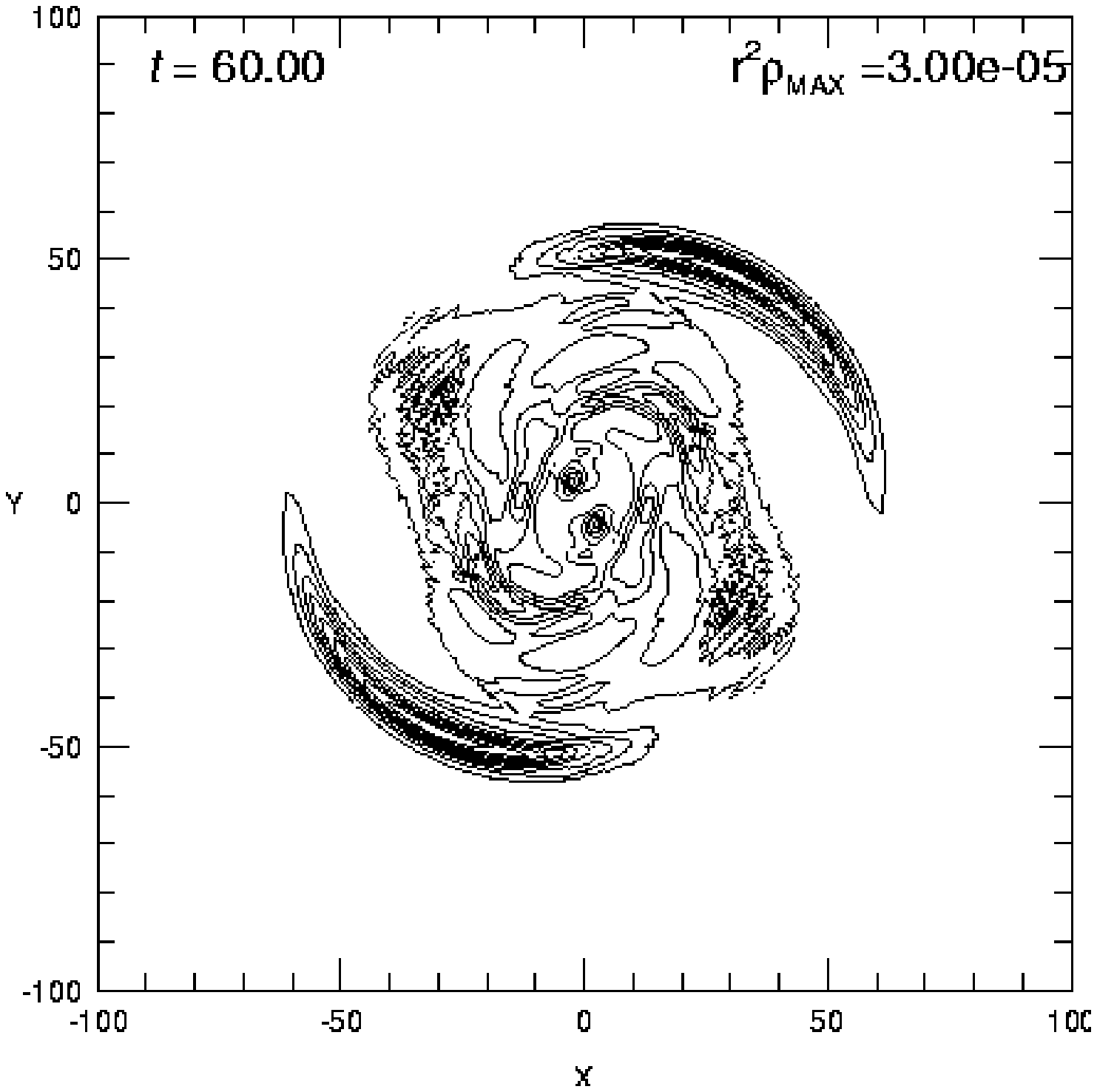}}
    \end{center}
  \end{minipage}
    \caption{$r^2 \dot{\tilde{\gamma}}_{ij}^2/ 32\pi$, which is
      expected to represent ``Energy density of the gravitational
      waves.''}
    \label{fig:bincol-gwden}
  \end{minipage}
\end{figure}

\begin{figure}[tbp]
  \begin{minipage}[t]{.48\textwidth}
    \begin{center}
    \leavevmode
    \scalebox{.29}{\includegraphics{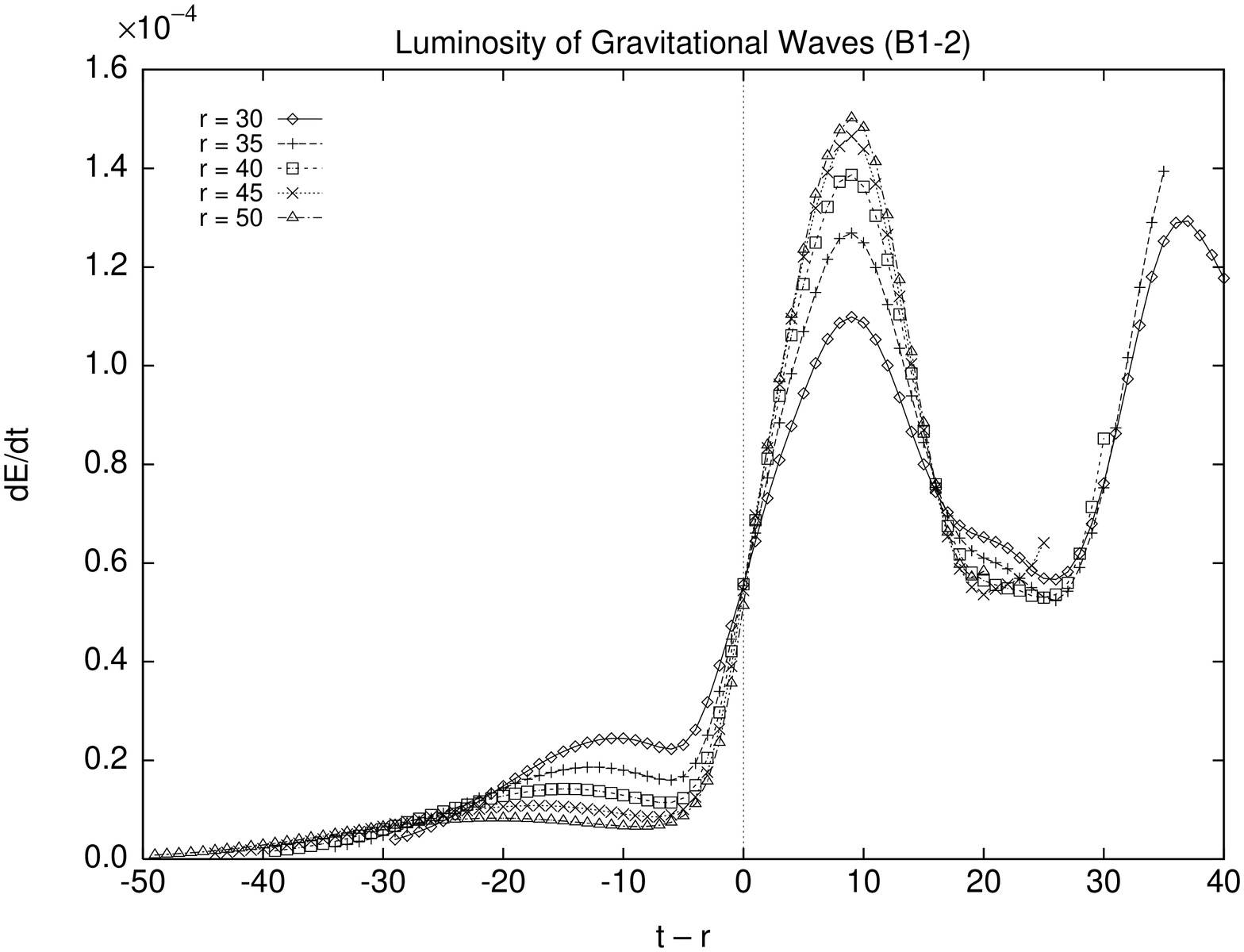}}
    \caption{Luminosity of gravitational waves $L_{\mbox{\scriptsize
          GW}}$ as a function of the retarded time $t - r$ calculated
      at $r = 30 \sim 50$.}
    \label{fig:bincol-flux}
    \end{center}
  \end{minipage}
  \hfill
  \begin{minipage}[t]{.48\textwidth}
  \begin{center}
    \leavevmode
    \scalebox{.29}{\includegraphics{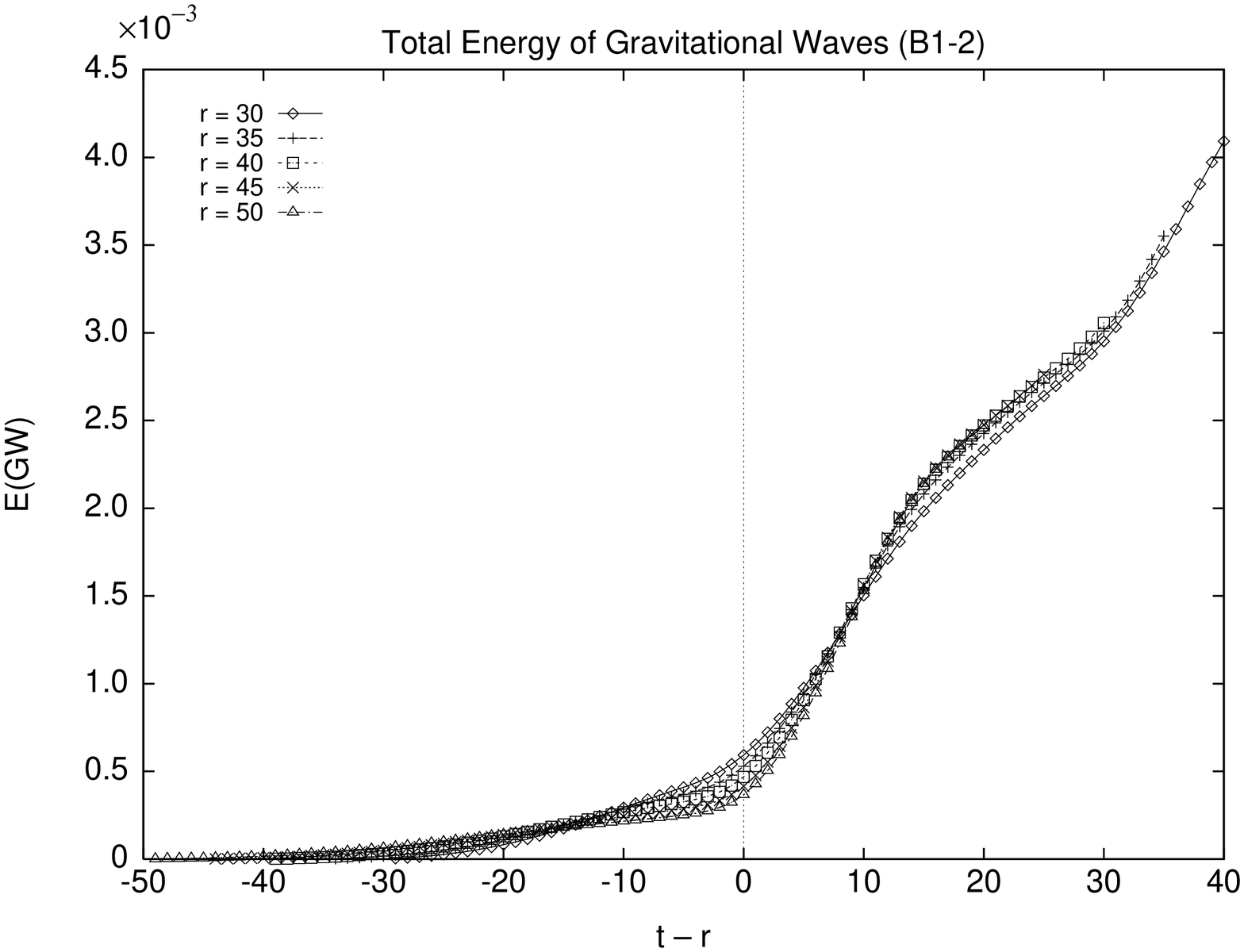}}
    \caption{Energy of gravitational waves $E_{\mbox{\scriptsize
          GW}}$ as a function of the retarded time $t - r$ calculated
      at $r = 30 \sim 50$.}
    \label{fig:bincol-gwen}
    \end{center}
  \end{minipage}
\end{figure}

We performed two simulations, BI1 and BI2, with different values of
$\Omega$. The total angular momentum divided by the square of the
gravitational mass are $0.99$ and $1.4$ and  for BI1 and  BI2,
respectively. Figure 6 shows the evolution of the density on the
$x$-$y$, $y$-$z$ and $x$-$z$ planes of BI2. The binary starts to
coalesce and rotates about 180 degree. In Fig.\ref{fig:bincol-gwden}
we show $flux\equiv r^2 \dot{\tilde{\gamma}}_{ij}^2/ 32\pi$ which can
be considered as the energy flux of the gravitational wave energy if
we take appropriate averaging. $flux$ shows an interesting feature of
generation and propagation of the waves. A spiral pattern appears on
the $x$-$y$ plane while different patterns with peaks around $z$-axis
appear on the $x$-$z$ and $y$-$z$ planes. This can be explained  by
the quadrupole wave pattern. In Fig.\ref{fig:bincol-flux} we show the
luminosity of the gravitational waves as a function of the retarded
time $t-r$ calculated at various $r$. For large $r$ it seems that the
luminosity converges, which suggests that the gravitational waves are
calculated correctly in our simulations.

The total energy of the gravitational radiation as a function of the
retarded time is shown in Fig.\ref{fig:bincol-gwen}.
The total energy amounts to
$4\times 10^{-3}M$(0.2\% of the total mass) and $1\times 10^{-3}M$
(0.05\% of the total mass) for BI2 and BI1, respectively.

In conclusion, the present test simulations suggest future fruitful
development of 3D fully general relativistic simulations of coalescing
binary neutron stars.

Numerical simulations were performed by  VPP300 at NAO. This work was
also supported by a Grant-in-Aid for Basic Research of the Ministry of
Education, Culture, and Sports No.08NP0801.

\end{document}